\documentclass[aps,pra,twocolumn,showpacs,superscriptaddress]{revtex4-1}
\usepackage{bm}
\usepackage{mathrsfs}
\usepackage{amsmath}
\usepackage{amssymb}
\usepackage{graphicx}
\usepackage{amsfonts}
\usepackage{amsthm}
\usepackage{color}
\usepackage{dcolumn}
\usepackage{txfonts}
\usepackage{epsfig}

\begin{document}

\title{Dissipation in the Generalized Gibbs Ensemble}
\author{Bo-Bo Wei}
\affiliation{School of Physics and Energy, Shenzhen University, 518060 Shenzhen, China}

\begin{abstract}
In this work, we show that the dissipation in a many-body system under an arbitrary non-equilibrium process is related to the R\'{e}nyi divergences between two states along the forward and reversed dynamics under very general family of initial conditions. This relation generalizes the links between dissipated work and Renyi divergences to quantum systems with conserved quantities whose equilibrium state is described by the generalized Gibbs ensemble. The relation is applicable for quantum systems with conserved quantities and can be applied to protocols driving the system between integrable and chaotic regimes. We demonstrate our ideas by considering the one-dimensional transverse quantum Ising model which is driven out of equilibrium by the instantaneous switching of the transverse magnetic field.
\end{abstract}
\maketitle

\section{Introduction}
The non-equilibrium dynamics of quantum many-body systems is one of the most fascinating branch in physics. The second law of thermodynamics told us that the work done under a non-adiabatic change of the parameter of the system is always greater than the free energy change. While in 1997, Jarzynski \cite{Jarzynski1997} made an milestone development in non-equilibrium thermodynamics by discovering that a classical system initialized in a thermodynamic equilibrium state the work done on the system under an arbitrary change of control parameters is related to the equilibrium free energy difference between the initial and the final thermal equilibrium states for the control parameters. The Jarzynski equality links the equilibrium free energy difference to the work done in an arbitrary driving process and therefore provides a novel method to
study thermodynamics in nanscale sytems through non-equilibrium measurement \cite{PNAS2001,Science2005,Nature2005,EPL2005,PT2005,PRL2006,PRL2007}. Jarzynski equality was extended to finite quantum mechanical systems \cite{arXiv2000a,arXiv2000b,PRL2003,Talker2007} and has been verified experimentally in an ion trap system \cite{Kim2015}. The discovery of the Jarzynski equality has led to extensive investigation of fluctuation relations in non-equlibrium thermodynamics \cite{RMP2009,Jar2011,RMP2011}.

Recently, the author and his collaborator found that \cite{Wei2017a} the dissipation under an arbitrary driving protocol is related to the R\'{e}nyi divergences between a microscopic
state in the forward process and a microscopic state in its time reversed process. Because the R\'{e}nyi divergence is a measure of distinguishability of two states \cite{Renyi1961,Erven2014,Beigi2013,Lennert2013}, the relation between dissipation and Renyi divergences \cite{Wei2017a} tells us that dissipation comes from breaking of microscopic reversibility. On the other hand, the relation between dissipation and Renyi divergences \cite{Wei2017a} tells us that we can measure the family of R\'{e}nyi divergences by non-equilibrium work measurement in a microscopic process \cite{Wei2017a}. The relation between dissipation and R\'{e}nyi divergences has been recently verified in a superconducting qubit system \cite{Dissipationexp2017}. Note that the maximum dissipated work was shown to be proportional to the infinite order R\'{e}nyi divergences in \cite{Halpern2015}. Recently the relation between dissipation and Renyi divergences was generalized to $\mathcal{PT}$ quantum systems with unbroken $\mathcal{PT}$ symmetry \cite{Wei2017b}.

One of the fundamental assumptions in deriving fluctuation relations in non-equilibrium thermodynamics \cite{RMP2011} is that the system is initialized in the equilibrium state described by Gibbs ensemble. However, the steady state of an important number of quantum systems can not be described by Gibbs ensemble \cite{GGEexp0}. For instance, the quantum integrable systems have a large number of conserved quantities which put constraints on the available phase space of a system and strongly affects the dynamics of the system \cite{GGE1,GGE2,GGE3,GGE4}.  Instead of relaxing to a steady state described by the usual thermodynamical ensembles, a generalized Gibbs ensemble (GGE) was proposed \cite{GGEtheory1,GGEreview2016} to describe the corresponding steady states and have been verified experimentally \cite{GGEexp1}. These results rises questions of how the impact of conserved quantities on the fluctuation relations when a system is driven out of equilibrium and more generally how do physical observables fluctuates when the number of conserved quantities varies during time evolution. Recently, the quantum Jarznski equality and Crooks relations have been derived when the initial state is described by GGE \cite{GGEfluctuation2017}. The motivation of the present work is to investigate the dissipation in driving quantum system which is initialized in the GGE state? We found that indeed in this case the dissipation is also determined by the R\'{e}nyi divergences. We then make use of a quantum integrable model, the transverse Ising model to theoretically verify our findings.

The remainder of this paper is organized as follows: In Sec.~II, we briefly review the formalism of generalized Gibbs ensemble. In Sec.~III, we first define the forward process and the reversed process for quantum systems with conserved quantities and then derive the relation between the dissipated work in driving quantum system which is initialized in the GGE and the R\'{e}nyi divergences between two quantum states. In Sec. IV, we use a simple quantum integral systems to analytically verify our finding. In Sec.~V, we make a brief summary of the results.

\section{Fundamentals of the Generalized Gibbs Enembles}
It is well known that the principle of entropy maximization \cite{Jaynes1957a,Jaynes1957b} leads to the famous thermodynamical ensembles, such as the canonical ensemble and the grand canonical ensemble,
which are constrained by conserved quantities such as energy and particle number. However, many physical systems may contain
additional conserved quantities and this raises the question of whether there exists a more general statistical description for the steady states
of quantum many-body systems. The quantum mechanical description of systems with more conserved quantities can also be built on information theory formulation of statistical mechanics \cite{Jaynes1957a,Jaynes1957b}. In this approach, the generalized Gibbs ensemble has been proposed to take account of the constraints of conserved quantities. In the GGE, the equilibrium state of a system with Hamiltonian $H$ is given by the density matrix \cite{GGEtheory1},
\begin{eqnarray}
\rho_{\text{GGE}}&=&\frac{1}{Z}\exp\left(-\beta H-\sum_{k=1}^M\beta_kI_k\right).
\end{eqnarray}
Here $Z(\beta,\beta_1,\cdots,\beta_M)=\text{Tr}\left[\exp\left(-\beta H-\sum_{k=1}^M\beta_kI_k\right)\right]$ is the partition function and $I_k,k=1,2,\cdots,M$ is the conserved quantities of the system, which commutes with the Hamiltonian of the system. We assume the conserved operators commute with each other so that we can measure them simultaneously. The generalized inverse temperatures $\{\beta,\beta_1,\cdots,\beta_M \}$ are fixed by requiring that averages over $\rho_{GGE}$ reproduce the known average values of the energy and conserved quantities,
\begin{eqnarray}
\langle H\rangle&=&\text{Tr}[\rho_{GGE}H]=\bar{E},\\
\langle I_k\rangle&=&\text{Tr}[\rho_{GGE}I_k]=\bar{I}_k.
\end{eqnarray}
The generalized Gibbs ensemble has been experimentally verified recently \cite{GGEexp1}. What is the dissipation for a quantum many-body system whose initial state is described by GGE being driving out of equilibrium? This is the main target of this work.

\section{Non-equilibrium Quantum thermodynamics}
Let us consider a finite quantum system with conserved quantities so that its steady state can be described by the generalized Gibbs ensemble. We are interested in a non-equilibrium process induced by a time-dependent Hamiltonian $\mathcal{H}(t)$ which is controlled by an external agent.

\subsection{Forward Driving Process}
Let us first describe the \emph{forward driving process}.\\
(1). The system is initialized in the equilibrium state corresponding to Hamiltonian $H=\mathcal{H}(0)$. If the Hamiltonian $H$ has a number of conserved quantities denoted by the operators $I_k,k=1,2,\cdots,M$, then the equilibrium state can be written in the form of the GGE with $M+1$ parameters, $\vec{\beta}=\{\beta,\beta_1,\beta_2,\cdots,\beta_M\}$,
\begin{eqnarray}
\rho_{F}(0)&=&\frac{e^{-\beta H-\sum_{k=1}^{M}\beta_{k}I_k}}{Z(\beta,\beta_1,\cdots,\beta_{M})}.
\end{eqnarray}
Here $Z(\beta,\beta_1,\cdots,\beta_k)=\text{Tr}[e^{-\beta H-\sum_{k=1}^{M}\beta_{k}I_k}]$ is the partition function of the initial equilibrium state.
We build a basis of the Hilbert space with eigenvectors $|\vec{i}\rangle=|i_0,i_1,\cdots,i_M,\eta\rangle$, where the quantum number $i_0$ characterizes the energy eigenvalue by the Schr\"{o}dinger equation $H|i_0\rangle=E_{i_0}|i_0\rangle$, $i_k$ labels the eigenvalues of the conserved quantitiets $I_k$ through $I_k|i_k\rangle=I_{k,i_k}|i_k\rangle$ and $\eta$ contains the additional quantum numbers required to completely determine a basis state.\\
(2). At time $t=0$, one performs the first projective measurement of $H$ and $I_k$ on the system, obtaining definite values for its energy $\mathcal{E}_{i_0}\in\{E_i\}$ and other observables, $\mathcal{I}_{k,i_k}\in\{I_{k,i}\}, k=1,2,\cdots,M$ with the corresponding probability,
\begin{eqnarray}
p_{\vec{i}}&=&\frac{e^{-\beta \mathcal{E}_{i_0}-\sum_{k=1}^M\beta_k\mathcal{I}_{i_k}}}{Z(\beta,\beta_1,\cdots,\beta_k)},
\end{eqnarray}
where $Z(\beta,\beta_1,\cdots,\beta_k)=\sum_{\vec{i}}e^{-\beta \mathcal{E}_{i_0}-\sum_{k=1}^M\beta_k\mathcal{I}_{i_k}}$ is the partition function of the initial equilibrium state. Simultaneously the initial state $\rho_F(0)$ is projected into the corresponding eigenstate $|\vec{i}\rangle$.\\
(3). The system is driven out of equilibrium by a time-dependent Hamiltonian $\mathcal{H}(t)$ in a time interval $0<t<\tau$. This defines a unitary time evolution operator $\mathcal{U}_{0,\tau}$ as the solution of the time-dependent Schr\"{o}dinger equation,
\begin{eqnarray}
i\partial_t\mathcal{U}_{0,t}=\mathcal{H}(t)\mathcal{U}_{0,t}.
\end{eqnarray}
The corresponding time development operator is
\begin{eqnarray}
\mathcal{U}_{0,\tau}&=&\mathcal{T}e^{-i\int_0^{\tau}dt\mathcal{H}(t)}.
\end{eqnarray}
Thus the state of the system at time $\tau$ becomes $\mathcal{U}_{0,\tau}|\vec{i}\rangle$.\\
(4). At time $t=\tau$, the second projective measurement is performed and the system is projected on the eigenstates of the final Hamiltonian, $H'=\mathcal{H}(\tau)$. In general, the conserved operators $I_k$ will not commute with $H(t)$ for $t>0$, so we identify the new set of conserved quantities, $\{I_l',l=1,2,\cdots,M'\}$, which commute with $H'$. Assuming that these operators commute with each other so that we can measure them simultaneously. This second projective measurement provides access to the quantities $\mathcal{E}_{f_0}'$ and $\mathcal{I}_{l,f_l}'$, each belonging to the spectrum of the corresponding operator defined in the analogy to the case at $t=0$ with the conditional probability,
\begin{eqnarray}
p_{\vec{i}\rightarrow\vec{f}}=\left|\langle \vec{f}|\mathcal{U}_{0,\tau}|\vec{i}\rangle\right|^2.
\end{eqnarray}
Here $|\vec{f}\rangle=|f_0,f_1,\cdots,f_{M'},\eta'\rangle$ is the common eigenstate of $H'$ and $\{I_l',l=1,2,\cdots,M'\}$. Thus at the end of a single realization of the forward process, one has collected the data set $\{\mathcal{E}_{i_0},\{\mathcal{I}_{k,i_k},k=1,2,\cdots,M\};\mathcal{E}_{f_0}',\mathcal{I}_{l,f_l}',l=1,2,\cdots,M'\}$ associated to the parameters of the initial GGE state. With these data set, the generalized work defined in the forward process is \cite{GGEfluctuation2017}
\begin{eqnarray}
\mathcal{W}=\left(\beta'\mathcal{E}_{f_0}'+\sum_{l=1}^{M'}\beta_l'\mathcal{I}_{l,f_l}'\right)-\left(\beta\mathcal{E}_{i_0}+\sum_{k=1}^{M}\beta_k\mathcal{I}_{k,i_k}\right).
\end{eqnarray}
To simplify the notation, we introduce the short hand notation,
\begin{eqnarray}
\mathcal{A}_{\vec{i}}&=&\beta\mathcal{E}_{i_0}+\sum_{k=1}^{M}\beta_k\mathcal{I}_{k,i_k},\\
\mathcal{A}_{\vec{f}}'&=&\beta'\mathcal{E}_{f_0}'+\sum_{l=1}^{M'}\beta_l'\mathcal{I}_{l,f_l}'.
\end{eqnarray}
With the state $\vec{i}=|i_0,i_1,\cdots,i_M,\eta\rangle$ and $\vec{f}=|f_0,f_1,\cdots,f_{M'},\eta'\rangle$. Then the generalized work distribution in the forward process is \cite{GGEfluctuation2017},
\begin{eqnarray}\label{QWD}
P_F(\mathcal{W})=\sum_{\vec{i},\vec{f}}p_{\vec{i}}\left|\langle \vec{f}|U|\vec{i}\rangle\right|^2\delta\left[\mathcal{W}-(\mathcal{A}'_{\vec{f}}-\mathcal{A}_{\vec{i}})\right].
\end{eqnarray}
The characteristic function of quantum work distribution, which is the Fourier transform of the quantum work distribution, is given by
\begin{eqnarray}
G(u)&=&\int_{-\infty}^{\infty}d\mathcal{W}P_F(\mathcal{W})e^{iu\mathcal{W}},\\
&=&Z^{-1}\text{Tr}\left[\mathcal{U}_{0,\tau}e^{-(1+iu)\left(\beta H+\sum_{k=1}^M\beta_kI_k\right)}\mathcal{U}_{0,\tau}^{\dagger}e^{iu\left(\beta'H'+\sum_{l=1}^{M'}\beta_l'I_l'\right)}\right].\label{CF}\nonumber\\
\end{eqnarray}
Note that the forward characteristic function in Equation \eqref{CF} has a similar expression to the Loschmidt echo or central spin decoherence \cite{Yang2017}, which has been shown to be deeply related to the partition function with a complex parameter \cite{Wei2012,Wei2014,Wei2015,Peng2015,LYExp2015,Wei2017a1,Wei2017b1,Wei2017b1} and could be experimentally measured \cite{Yang2017}.

\subsection{Reversed Driving Process}
Now let us define the reversed process in quantum system. \\
(1). In the reversed process, we begin with initializing the system in the time reversed state of the equilibrium state corresponding to Hamiltonian $H'=\mathcal{H}(\tau)$,
\begin{eqnarray}
\rho_{R}(0)&=&\Theta \frac{e^{-\beta'H'-\sum_{l'=1}^{M'}\beta_{l'}I_l'}}{Z(\beta',\beta_1',\cdots,\beta_{M'}')}\Theta^{-1}.
\end{eqnarray}
Here $\Theta$ is the time reversal operator. The equilibrium state $\rho_{R}(0)$ is the form of GGE with $M'+1$ parameters, $\vec{\beta}'=\{\beta',\beta_1',\beta_2',\cdots,\beta_{M'}'\}$.\\
(2). At time $t=0$,  we perform the first projective measurement $\Theta H'\Theta^{-1}$ and $\Theta I_l'\Theta^{-1}$ on the system and the state is projected on the basis $\Theta|\vec{f}\rangle$ of simultaneous eigenvectors of $\Theta H'\Theta^{-1}$ and $\Theta I_l'\Theta^{-1}$, obtaining definite values for its energy $\mathcal{E}_{f_0}'$ and $\mathcal{I}_{l,f_l}',i=1,2,\cdots,M'$.\\
(3). In the third step, the system is driven out of equilibrium by the time reversed Hamiltonian \cite{Stra1994}
\begin{eqnarray}
\mathcal{H}_R(t)&=&\Theta\mathcal{H}(\tau-t)\Theta^{-1},
\end{eqnarray}
for a time duration $\tau$. The corresponding time evolution operator in the reversed process is
\begin{eqnarray}
\mathcal{V}_{0,\tau}&=&\mathcal{T}e^{-i\int_0^{\tau}dt\Theta\mathcal{H}(\tau-t)\Theta^{-1}}.
\end{eqnarray}
So the density operator in the reversed process at time $t$ is
\begin{eqnarray}\label{rs}
\rho_R(t)&=&\mathcal{V}_{0,t}\rho_R(0)\mathcal{V}_{0,t}^{\dagger}.
\end{eqnarray}
(4). At time $t=\tau$, the second projective measurement of $\Theta H \Theta^{-1}$ and $\Theta I_k\Theta^{-1}$ is performed and yielding $\mathcal{E}_{i_0}$ and $\mathcal{I}_{k,i_k}$ for the energy and other conserved quantities. Thus a single realization of the reversed process provides the data set $\{\mathcal{E}_{f_0}',\{\mathcal{I}_{l,f_l}',l=1,2,\cdots,M'\};\mathcal{E}_{i_0},\{\mathcal{I}_{k,i_k},k=1,2,\cdots,M\}\}$ associated to the parameters $\vec{\beta}'$ of the corresponding initial state.

\subsection{Dissipation in the Generalized Gibbs Ensemble}
To calculate dissipation in the GGE, we shall make use of a Lemma \cite{Wei2017a}\\
\textbf{Lemma}. A density operator $\rho_F(t)$ in the forward driving process and the density operator $\rho_R(t)$ in the time reversed process satisfy, for any finite real numbers $a,b\in\mathcal{R}$,
\begin{eqnarray}
\text{Tr}[\left(\Theta^{-1}\rho_R(\tau-t)\Theta\right)^a\rho_F(t)^b]=\text{Tr}[\left(\Theta^{-1}\rho_R(0)\Theta\right)^a\rho_F(\tau)^b].
\end{eqnarray}
This Lemma is a consequence of time reversal symmetry and unitarity of the time development and was proved in \cite{Wei2017a}. Here we just make use of this Lemma in the following derivation.
From the definition of quantum work distribution, Equation\eqref{QWD}, we have
\begin{eqnarray}
&&\langle\Big( e^{-\mathcal{W}}\Big)^z\rangle_F\nonumber\\
&=&\int d\mathcal{W} P_F(\mathcal{W})e^{-z\mathcal{W}},\\
&=&\sum_{\vec{i},\vec{f}}p_{\vec{i}}\left|\langle \vec{f}|\mathcal{U}_{0,\tau}|\vec{i}\rangle\right|^2e^{-z\mathcal{A}'_{\vec{f}}+z\mathcal{A}_{\vec{i}}},\\
&=&Z^{-1}\sum_{\vec{i},\vec{f}}\left|\langle \vec{f}|\mathcal{U}_{0,\tau}|\vec{i}\rangle\right|^2e^{-z\mathcal{A}'_{\vec{f}}}e^{-(1-z)\mathcal{A}_{\vec{i}}},\\
&=&Z^{-1}\sum_{\vec{i},\vec{f}}\langle\vec{f}|\mathcal{U}_{0,\tau}|\vec{i}\rangle\langle \vec{i}|\mathcal{U}_{0,\tau}^{\dagger}|\vec{f}\rangle e^{-z\mathcal{A}'_{\vec{f}}}e^{-(1-z)\mathcal{A}_{\vec{i}}},\\
&=&Z^{-1}\sum_{\vec{i},\vec{f}}\langle\vec{f}|\mathcal{U}_{0,\tau}e^{-(1-z)(\beta H+\sum_k\beta_kI_k)}|\vec{i}\rangle\langle \vec{i}|\mathcal{U}_{0,\tau}^{\dagger}e^{-z(\beta' H'+\sum_l\beta_l'I_l')}|\vec{f}\rangle,\nonumber\\
&=&Z^{-1}\sum_{\vec{f}}\langle\vec{f}|\left(\mathcal{U}_{0,\tau}e^{-(\beta H+\sum_k\beta_kI_k)}\mathcal{U}_{0,\tau}^{\dagger}\right)^{1-z}\left(e^{-(\beta' H'+\sum_l\beta_l'I_l')}\right)^z|\vec{f}\rangle,\nonumber\\
&=&\left(\frac{Z'}{Z}\right)^z\text{Tr}\left[\left(\mathcal{U}_{0,\tau}\rho_{F}(0)\mathcal{U}_{0,\tau}^{\dagger}\right)^{1-z}\left(\Theta^{-1}\rho_R(0)\Theta\right)^z\right],\\
&=&\left(\frac{Z'}{Z}\right)^z\text{Tr}\left[\left(\rho_{F}(\tau)\right)^{1-z}\left(\Theta^{-1}\rho_R(0)\Theta\right)^z\right],\label{d1}\\
&=&\left(\frac{Z'}{Z}\right)^z\text{Tr}\left[\left(\rho_{F}(t)\right)^{1-z}\left(\Theta^{-1}\rho_R(\tau-t)\Theta\right)^z\right],\label{d2}\\
&=&e^{-\Delta F}e^{(z-1)S_z\left(\Theta^{-1}\rho_R(\tau-t)\Theta||\rho_F(t)\right)}.\label{d3}
\end{eqnarray}
Here $z$ is a finite real number and $\Delta F\equiv F(\beta',\beta_1',\cdots,\beta_{M'}')-F(\beta,\beta_1,\cdots,\beta_M)$ is the free energy difference between the equilibrium states at the initial and final control parameters of the GGE, where the free energy is related to the partition function by $F=-\ln Z$. From Equation \eqref{d1} to Equation \eqref{d2}, we have made use of the Lemma list above. In the last step, we have made use of definition of the order-$z$ quantum R\'{e}nyi divergences between two density matrices $\rho_1$ and $\rho_2$ \cite{Beigi2013,Lennert2013}, $S_z(\rho_1||\rho_2)\equiv\frac{1}{z-1}\ln[\text{Tr}[\rho_1^{z}\rho_2^{1-z}]]$, which is information theoretic generalization of standard relative entropy \cite{Renyi1961}. If we identify $\mathcal{W}-\Delta F$ as the dissipated work $\mathcal{W}_{\text{diss}}$ in the non-equilibrium process, then the fluctuation of the dissipated work is given by,
\begin{eqnarray}\label{central0}
\langle\Big( e^{-\mathcal{W}_{\text{diss}}}\Big)^z\rangle&=&\exp\left[(z-1)S_z\left(\Theta^{-1}\rho_R(\tau-t)\Theta||\rho(t)\right)\right].
\end{eqnarray}
Now we make some remarks on the above relation:\\
(1).~In Equation \eqref{central0}, $z$ is a free parameter and can take any positive real numbers. If we set $z=1$, Equation \eqref{central0} recovers the Jarzynski equality in the generalized Gibbs ensemble \cite{GGEfluctuation2017}.\\
(2).~Fluctuation of the dissipated work is independent of time $t$ because the density operators $\rho_F(t)$ and $\rho_R(\tau-t)$ appeared on the right hand side of Equation \eqref{central0} may be evaluated at any intermediate time $t\in[0,\tau]$, which is a consequence of the time reversal symmetry and unitarity of quantum dynamics. \\
(3).~The quantum R\'{e}nyi divergences can measure distinguishability of two states \cite{Erven2014,Beigi2013} and thus Equation \eqref{central0} tells us that dissipation comes from the breaking of time reversal symmetry between the forward and its time reversed dynamics.
\\(4).~Equation \eqref{central0} connects the fluctuations of the dissipated work to the quantum R\'{e}nyi divergences between two non-equilibrium quantum states along the forward process and its time reversed process. It is clear that the dissipated work is a macroscopic quantity while on the other hand the R\'{e}nyi divergences between two quantum states is a microscopic quantity. Thus the relation connects a microscopic quantity to a macroscopic quantity in non-equilibrium thermodynamics. Various moments of the dissipated work are given by
\begin{eqnarray}
\langle \mathcal{W}_{\text{diss}}^n\rangle&=&\text{Tr}\Big[\rho_F(t)\mathcal{T}_n\Big(\ln[\rho_F(t)]-\ln[\Theta^{-1}\rho_R(\tau-t)\Theta]\Big)^n\Big],\label{momentsquan}
\nonumber \\
\end{eqnarray}
where $n=1,2,3,\cdots$ and $\mathcal{T}_n$ is an ordering operator which sorts that in each term of the binomial expansion of $\Big(\ln[\rho_F(t)]-\ln[\Theta^{-1}\rho_R(\tau-t)\Theta]\Big)^n$, $\ln[\rho_F(t)]$ sits on the left of $\ln[\Theta^{-1}\rho_R(\tau-t)\Theta]$. In particular for $n=1$, it is
\begin{eqnarray}\label{ex}
\langle \mathcal{W}_{\text{diss}}\rangle&=&D[\rho_F(t)||\Theta^{-1}\rho_R(\tau-t)\Theta],\label{qdiss}
\end{eqnarray}
where $D[\rho_1||\rho_2]\equiv \text{Tr}[\rho_1(\ln\rho_1-\ln\rho_2)]$ is the von Neumann relative entropy \cite{relativeentropy} between two density matrices. Note that it has been already known that the mean value of dissipated work is related to the relative entropy between the forward dynamics and its time reversed dynamics for non-equilibrium dynamics in the Gibbs ensemble \cite{Kawai2007,Jarzynski2006,Jarzynski2009,Parrondo2009,Deffner2010,Vedral2012,Serra2015}.
Our equation \eqref{ex} generalizes the the relation between dissipated work and the relative entropy to systems with very general initial conditions. \\
(5). Equation \eqref{central0} establishes an exact relation between the generating function of the dissipated work and the quantum R\'{e}nyi divergences between two non-equilibrium states and therefore we can measure the arbitrary ordre of R\'{e}nyi divergences from the non-equilibrium work measurement in a closed quantum system driving out of equilibrium. Because the characteristic function of quantum work distribution can be measured from the Ramsey interference of a single spin \cite{Dorner2013,Mazzola2013,measure1}, one can also measure the R\'{e}nyi divergences between two quantum states from Ramsey interference experiment.\\
(6). If the Hamiltonian $H$ and $H'$ have no conserved quantities except the energy, namely the number of conserved quantities for the initial and final Hamiltonian $M=M'=0$, and the forward and reversed processes begin at the same inverse temperature $\beta=\beta'$, then $\mathcal{W}=\beta(\mathcal{E}_{f_0}'-\mathcal{E}_{i_0})=\beta W$ where the ordinary work $W$ is the energy difference between the initial and final states. Thus Equation \eqref{central0} recovers the case for the Gibbs ensemble \cite{Wei2017a}.

\section{Physical Model Study}
The transverse Ising model in one-dimension (1D) is an integrable quantum system and we make use of this simple model to theoretically verifying the relation between dissipated work and the R\'{e}nyi divergences in the GGE. The Hamiltonian of the 1D quantum Ising model is written as
\begin{eqnarray}
H=-J\sum_{j=1}^N\Big[\sigma_j^x\sigma_{j+1}^x+\lambda\sigma_j^z\Big],
\end{eqnarray}
where $J$ is the coupling strength between nearest neighbor Ising spins along $x$ direction (we take $J=1$), $\lambda$ is a dimensionless parameter quantifying the strength of magnetic field along $z$ direction, $\sigma_j^{\alpha}(\alpha=x,y,z)$ is the Pauli operator of the $j$-th spin along $\alpha=x,y,z$ direction and we impose periodic boundary conditions for Pauli spins by requiring that $\vec{\sigma}_{N+1}=\vec{\sigma}_1$.

The 1D quantum Ising model may be exactly diagonalized through
three successive transformations, namely, the Jordan-Wigner
transformation, Fourier transformation and Bogoliubov
transformation \cite{Lieb1961,QPT2011}. After these three transformations, the Hamiltonian of 1D quantum Ising model finally becomes \cite{QPT2011}
\begin{eqnarray}
H(\lambda)&=&\sum_k\epsilon_k(\lambda)b_k^{\dagger}b_k=\sum_k\epsilon_kI_k.
\end{eqnarray}
Here $b_k$ and $b_k^{\dagger}$ are the fermion annihilation and creation operators of momentum $k$, $\epsilon_k(\lambda)=2\sqrt{1+\lambda^2-2\lambda\cos k}$ is the excitation energy for creating a fermion with momentum $k$, $I_k=b_k^{\dagger}b_k$ is the Fermionic number operators and the Fermionic number operators are all conserved quantities of $H(\lambda)$. Note that the set of conserved quantities $I_k$ are different if $\lambda$ changes.

We now consider a sudden quench process in the 1D quantum Ising model where the transverse magnetic field is tuned from $\lambda_i$ to $\lambda_f$ very quickly. At $t=0$, the quantum Ising model is prepared in a thermal equilibrium state described by GGE with density matrix $\rho_F(0)=e^{-\sum_k(\beta+\beta_k) I_k}/Z(\beta,\beta_1,\cdots,\beta_N)$ with $Z(\beta,\beta_1,\cdots,\beta_N)=\text{Tr}[e^{-\sum_k(\beta+\beta_k) I_k}]$ is the partition function in the GGE and $N$ is the total number of spins in the Ising model. Then we suddenly change the transverse magnetic field from $\lambda_i$ to $\lambda_f$. The characteristic function of work distribution in this sudden quench process can be calculated as
\begin{eqnarray}
G(u)&=&\int_{-\infty}^{\infty}dWP_F(\mathcal{W})e^{iu\mathcal{W}},\\
&=&Z^{-1}\text{Tr}\left[e^{-(1+iu)\left(\beta H+\sum_{k=1}^N\beta_kI_k\right)}e^{iu\left(\beta'H'+\sum_{l=1}^{N}\beta_l'I_l'\right)}\right],\\
&=&Z^{-1}\text{Tr}\left[e^{-(1+iu)\sum_{k=1}^N(\beta+\beta_k)I_k}e^{iu\sum_{k=1}^N(\beta'+\beta_k')I_k'}\right],\\
&=&\prod_{k>0}\frac{1+\frac{1+\cos\alpha_k}{2}\cos(A_+(u))+\frac{1-\cos\alpha_k}{2}\cos(A_-(u))}{1+\cosh(\beta+\beta_k)}.
\end{eqnarray}
Here $A_{\pm}(u)=(u-i)(\beta+\beta_k)\pm u(\beta'+\beta_k')$, $\alpha_k\equiv\theta_k^f-\theta_k^i$, $\theta_k^{i}$ and $\theta_k^f$ are shorthand notation for $\theta_k(\lambda_i)$ and $\theta_k(\lambda_f)$ with $\theta_k(\lambda_i)$ being defined by $\sin[2\theta_k(\lambda)]\equiv 2\sin k/\epsilon_k(\lambda)$ and $\cos[2\theta_k(\lambda)]\equiv 2(\lambda-\cos k)/\epsilon_k(\lambda)$.  In the sudden quench process, the free energy change is given by
\begin{eqnarray}
e^{-\Delta F}=\frac{Z'}{Z}=\prod_{k>0}\frac{[1+\cosh(\beta'+\beta_k')]}{[1+\cosh(\beta+\beta_k)]}.
\end{eqnarray}
On the other hand, the R\'{e}nyi divergences $S_z\left(\Theta^{-1}\rho_R(0)\Theta||\rho_F(\tau)\right)$ is
\begin{eqnarray}
&&\exp\left[(z-1)S_z\left(\Theta^{-1}\rho_R(0)\Theta||\rho_F(0)\right)\right],\nonumber\\
&=&Z^{-z}Z'^{z-1}\text{Tr}\left[e^{-z\sum_{k=1}^M(\beta+\beta_k)I_k}e^{-(1-z)\sum_{k=1}^M(\beta'+\beta_k')I_k'}\right],\\
&=&\prod_{k>0}\frac{1+\frac{1+\cos\alpha_k}{2}\cosh(B_+(z))+\frac{1-\cos\alpha_k}{2}\cosh(B_-(z))}{[1+\cosh(\beta+\beta_k)]^{1-z}[1+\cosh(\beta'+\beta_k')]^{z}}.
\end{eqnarray}
Here $B_{\pm}(z)=(z-1)(\beta+\beta_k)\pm z(\beta'+\beta_k')$.
Then it is easy to check that in the sudden quench process of the quantum Ising model, the dissipated work and R\'{e}nyi divergences between the initial state and the final state satisfy
\begin{eqnarray}
\langle \Big(e^{-\beta \mathcal{W}}\Big)^z\rangle&=&e^{-z\Delta F}e^{(z-1)S_z\left(\Theta^{-1}\rho_R(0)\Theta||\rho_F(0)\right)}.
\end{eqnarray}
Therefore we have analytically verified that the dissipation in the GGE is related to the R\'{e}nyi divergences of two states in an integrable quantum many-body system.

\section{Conclusions}
We have established that the dissipation in a quantum many-body system which experiences an arbitrary non-equilibrium process is linked to the R\'{e}nyi divergences between two states along the forward and reversed dynamics under very general family of initial conditions. This relation is a consequence of time reversal symmetry and unitarity of the dynamics and generalizes the connections between dissipated work and Renyi divergences to quantum systems with conserved quantities whose equilibrium state is described by the generalized Gibbs ensemble. The relation is applicable for quantum systems with conserved quantities and can be applied to protocols driving the system between integrable and chaotic regimes. We demonstrated our ideas analytically by studying the one-dimensional transverse quantum Ising model driving out of equilibrium by the instantaneous switching of the transverse magnetic field.

\begin{acknowledgements}
This work was supported by National Natural Science Foundation of China (Grants No. 11604220) and the Start Up Fund of Shenzhen University (Grants No. 2016018).
\end{acknowledgements}

\end{document}